# Dual-wavelength domain wall solitons in a fiber ring laser

**H. Zhang, D. Y. Tang*, L. M. Zhao and X. Wu**

[1]*School of Electrical and Electronic Engineering, Nanyang Technological University, Singapore 639798*

[*]*Corresponding author: edytang@ntu.edu.sg*

We report on the experimental observation of a new type of dark soliton in a fiber laser made of all normal group velocity dispersion fibers. It was shown that the soliton is formed due to the cross coupling between two different wavelength laser beams and has the characteristic of separating the two different wavelength laser emissions. Moreover, we show experimentally that the dual-wavelength dark solitons have a much lower pump threshold than that of the nonlinear Schrödinger equation dark solitons formed in the same laser.

Soliton formation in single mode fibers (SMFs) is a well-known effect and has been extensively investigated. It is now well recognized that the dynamics of the formed solitons is governed by the nonlinear Schrödinger equation (NLSE), and bright solitons are formed in the anomalous group velocity dispersion (GVD) fibers, while dark solitons are formed in the normal GVD fibers [1-3]. A fiber laser is mainly made of SMFs. It is natural to anticipate that under appropriate conditions solitary waves could be formed in the single mode fiber lasers. Indeed, both the bright and dark NLSE solitons have been experimentally observed in fiber lasers [4, 5].

In addition to the NLSE solitons, recently a novel new type of optical solitary waves known as the polarization domain wall solitons (PDWSs) were also experimentally revealed in fiber lasers [6]. Formation of PDWSs was first theoretically predicted by Haelterman and Sheppard [7]. It was shown that the cross coupling between the two orthogonal polarization components of light propagating in a dispersive Kerr medium could lead to the formation of a stable localized structure that separates domains of the two orthogonal polarization fields. Cross coupling between waves is a common phenomenon in a wide range of nonlinear physical systems. The experimental confirmation of PDWSs suggests that similar domain wall solitons could also be observed in other nonlinear wave coupling systems [8, 9]. In this letter, we report on the experimental observation of a dual-wavelength optical domain wall soliton (DWS) in a fiber ring laser made of all-normal GVD fibers. We show both experimentally and numerically that strong coupling between two different wavelength beams in the fiber laser can result in the formation of DWSs, representing as a stable dark intensity pulse separating the two different wavelength laser emissions.

We used a fiber laser whose cavity is as shown in [7]. Briefly, the cavity is made of ~ 5.0 m Erbium-doped fiber (EDF) with a GVD parameter of -32 (ps/nm)/km, ~ 6.1 m dispersion compensation fiber (DCF) with a GVD parameter of -2 (ps/nm)/km. A polarization sensitive isolator was employed in the cavity to force the unidirectional operation of the ring cavity, and an in-line polarization controller (PC) was used to fine-tune the linear cavity birefringence. A 10% fiber coupler was used to output the signal. The laser was pumped by a high power Fiber Raman Laser source of wavelength 1480 nm. The laser output was monitored with a 2 GHz photo-detector and displayed on a multi-channeled oscilloscope.

A major difference of the current laser to that of [5] is that in setting up the laser we have intentionally imposed large birefringence into the cavity. Consequently, the birefringence induced multi-pass filter effect becomes very strong [10] and can no longer be ignored for the laser. Under the combined action of the laser gain and the birefringent filter, the laser is forced to operate in a dual wavelength CW emission mode. Fig. 1(a) shows a typical spectrum of the laser under dual-wavelength emission. In our experiment the strength of the laser emission increased with the pump power. As the intensity of the laser emission was continuously increased, it was observed that a dark intensity pulse as shown in Fig. 1(b) could appear on the total laser output, and the stronger the pump power, the narrower became the dark pulse, as shown in Fig. 1(c). Tuning the orientation of the intra cavity PC, the bandpass wavelengths of the birefringence filter can be shifted. Therefore, the relative strength of the two laser emissions could be altered. Eventually single wavelength emission could also be obtained. It was found that under single wavelength operation no above dark pulses could be observed, indicating that the appearance of the dark pulse was related to the coupling of the two different wavelength beams.

Careful examining the laser output trace shown in Fig. 1(b) it came to our attention that the laser emission exhibited two distinctive intensity levels. We could experimentally identify that each of the intensity levels was related to the laser emission at one of the two wavelengths. Namely, the laser was not simultaneously emitting at the two wavelengths but alternating between them. The dark pulses always appeared at the position where the laser emission switched from one wavelength to the other. The dark pulses were very stable, showing that they are a DWS.

Under even stronger pump power, a state as shown in Fig. 2 was also obtained, where a new type of dark pulses with much narrower pulse width suddenly appeared. These new dark pulses moved with respect to the DWS. They not only have different darkness but also appeared randomly. We had studied previously the NLSE dark solitons in a fiber laser [5]. The observed features of the new dark pulses suggest that they are the NLSE dark solitons. In our experiment we could control the strength of the laser emission of either wavelength through shifting the filter frequencies. In this way we can suppress the appearance of the NLSE dark solitons on either of the two wavelength laser emissions. Fig. 3 shows a comparison of the laser emissions where one of the two-wavelength laser emissions is beyond the NLSE soliton threshold. Fig. 3(a) shows the laser emission spectra. The upper (lower) oscilloscope trace shown in Fig. 3(b) corresponds to the spectrum whose longer (shorter) wavelength spectral line has become broadened. Associated with the appearance of the NLSE dark solitons the corresponding laser emission spectrum became further broadened. The experimental results shown in Fig. 2 and Fig. 3 clearly demonstrated that the appearance of the DWSs has much lower pump threshold than the NLSE dark solitons.

To better understand the dual-wavelength DWS formation in our laser, we have further numerically simulated the operation of our laser under two wavelength emissions. The following coupled Ginzburg-Landau equations were used to describe the light propagation in the cavity fibers:

$$\frac{\partial u_1}{\partial z} = i\beta u_1 - \delta\frac{\partial u_1}{\partial t} - \frac{ik''}{2}\frac{\partial^2 u_1}{\partial t^2} + \frac{k'''}{6}\frac{\partial^3 u_1}{\partial t^3} + i\gamma\left(|u_1|^2 + 2|u_2|^2\right)u + \frac{g}{2}u_1 + \frac{g}{2\Omega_g^2}\frac{\partial^2 u_1}{\partial t^2},$$
$$\frac{\partial u_2}{\partial z} = -i\beta u_2 + \delta\frac{\partial u_2}{\partial t} - \frac{ik''}{2}\frac{\partial^2 u_2}{\partial t^2} + \frac{k'''}{6}\frac{\partial^3 u_2}{\partial t^3} + i\gamma\left(|u_2|^2 + 2|u_1|^2\right)v + \frac{g}{2}u_2 + \frac{g}{2\Omega_g^2}\frac{\partial^2 u_2}{\partial t^2} \qquad (1)$$

where $u_1$ and $u_2$ are the normalized envelopes of the optical pulses along the same polarization in the optical fiber but having different central wavelengths $\lambda_1$ and $\lambda_2$. $\beta = 2\pi\Delta n/(\lambda_1+\lambda_2)$ is the wave-number difference between the two optical waves. $\delta = \beta(\lambda_1+\lambda_2)/4\pi c$ is the inverse group velocity difference. $k''$ is the second order dispersion coefficient, $k'''$ is the third order dispersion coefficient and $\gamma$ represents the nonlinearity of the fiber. $g$ is the saturable gain coefficient of the fiber and $\Omega_g$ is the bandwidth of the laser gain. For undoped fibers $g$=0; for erbium doped fiber, we considered its gain saturation as

$$g = G\exp\left[-\frac{\int(|u|^2+|v|^2)dt}{P_{sat}}\right] \qquad (2)$$

Moreover, we considered the cavity feedback effect by circulating the light in the cavity [11]. To make the simulations possibly close to the experimental situation, we have assumed that the two wavelength beams have the same group velocity and used the following parameters: $\gamma$=3 $W^{-1}km^{-1}$, $\Omega_g$=16 nm, $k''_{DCF}$= 2.6 $ps^2/km$, $k''_{EDF}$= 41.6 $ps^2/km$, $k'''$= -0.13 $ps^3/km$, $P_{sat}$=500 pJ, cavity length L= 11.1 m, and G=120 $km^{-1}$.

A weak dual-wavelength beam with ~ 2.6 nm wavelength separation and an intensity switching between the two wavelengths was used as the initial condition. We let the light circulate in the cavity until a stable state is obtained. The CGLEs were solved using the split-step method. We found numerically that a stable DWS separating the laser emissions of different wavelengths could indeed be formed in our laser, as shown in Fig. 4. Fig. 4(a) shows the domain walls and the corresponding dark DWS calculated. Fig. 4(b) shows the optical spectrum of the laser emission.

We note that dual-wavelength DWS in a SMF was theoretically studied by Haelterman and Badolo [12]. To the best of our knowledge, no dual-wavelength optical DWSs have been experimentally confirmed. Based on our numerical simulations, we noticed that for the formation of the DWS an initial intensity alternation between the two wavelengths is crucial. We believe that in our laser the gain competition between the two laser beams could have played an important role on forming such an initial condition. It had been shown experimentally previously that gain competition could cause antiphase dynamics between two different wavelength beams in a fiber laser [13].

In conclusion, we have experimentally observed a new type of dark soliton in an erbium-doped fiber laser made of all-normal GVD fibers. It was shown that the formation of the soliton was a result of the mutual coupling between two different wavelength beams and the formed soliton has the characteristic of separating the two different wavelength laser emissions. It is an optical DWS. In addition, our experimental result has shown that the appearance of DWS has a much lower pump threshold than the NLSE dark solitons, and under strong nonlinear coupling the dual wavelength emission fiber laser is not simultaneously emitting two wavelengths by alternating between the two wavelengths.

## Figure captions:

Fig. 1: (a) Spectrum; (b) oscilloscope traces of the dual-wavelength optical domain wall. (c): the wall duration as a function of the pump strength.

Fig. 2: (a) Spectrum and (b) oscilloscope trace of the single dark soliton obtained through the paddles of PC. (For comparison)

Fig. 3: Generation of the multiple dark solitons at each individual wavelength of the optical domain wall. (a) Spectra and (b) oscilloscope traces, upper (lower) trace corresponds to spectral broadening at longer (shorter) wavelength.

Fig. 4: dual wavelength domain wall numerically calculated. (a) Domain wall profiles at particular roundtrip (b) Its corresponding spectra.

(a)

(b)

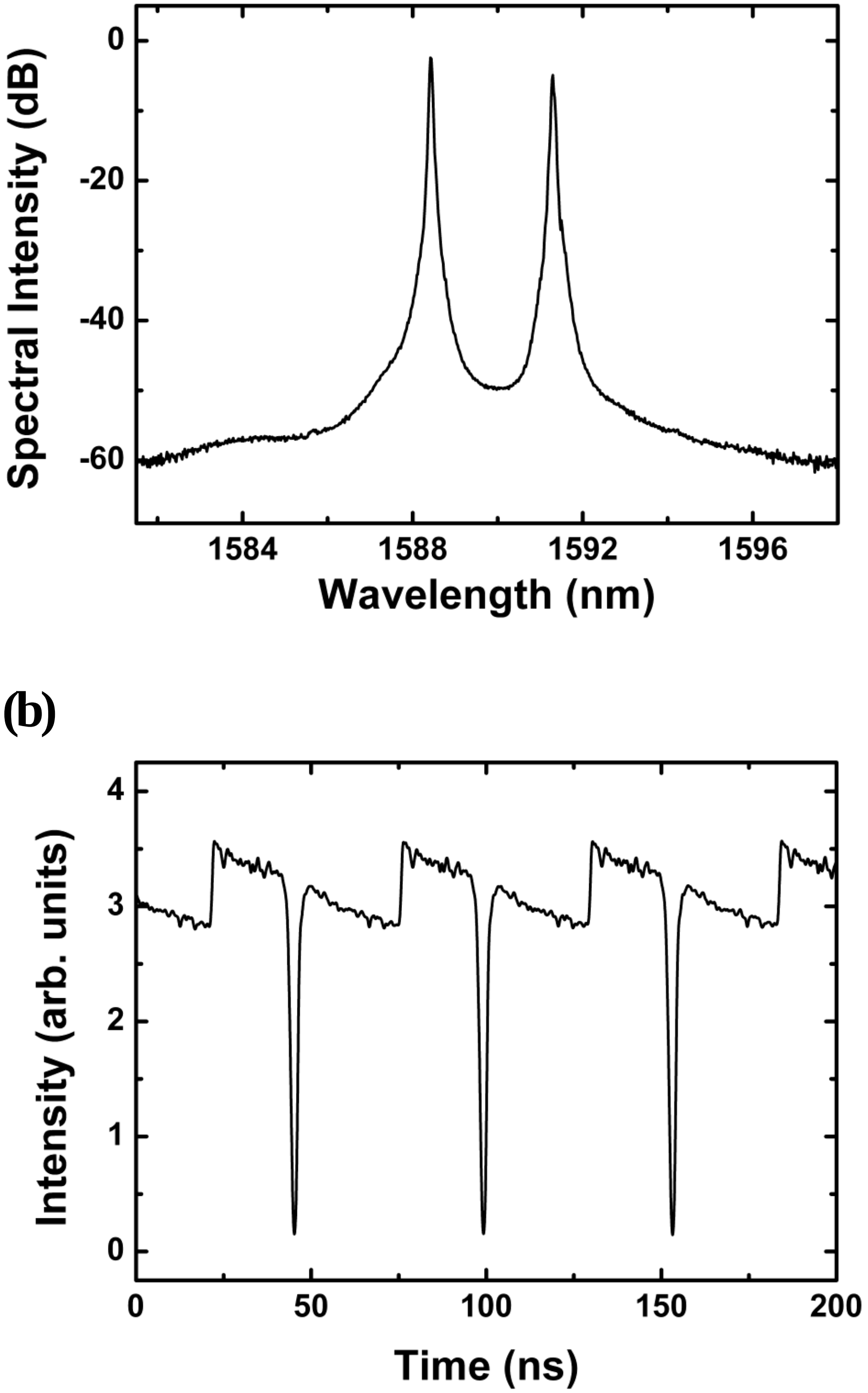

Fig.1 H. Zhang et. al.

(c)

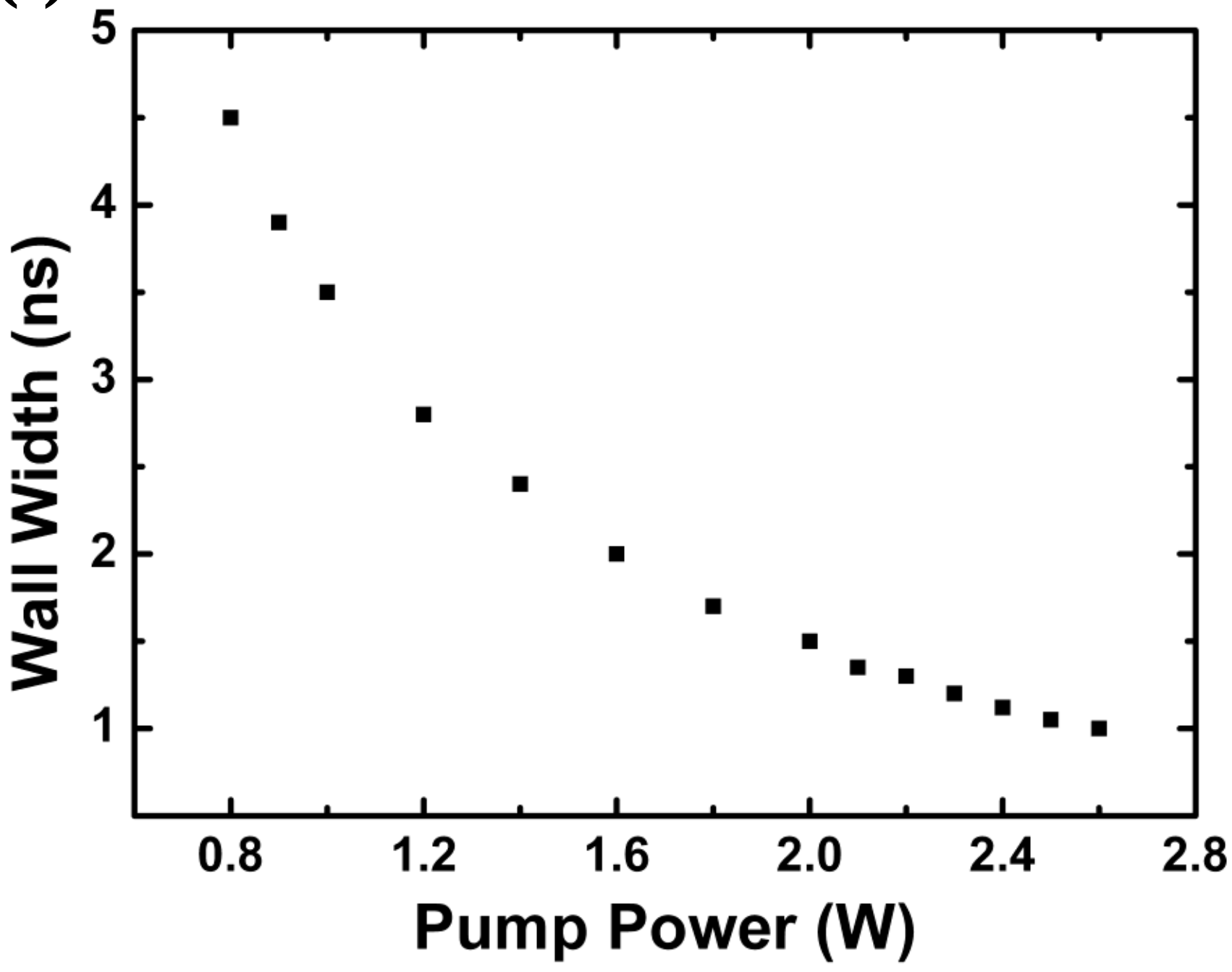

Fig.1 H. Zhang et. al.

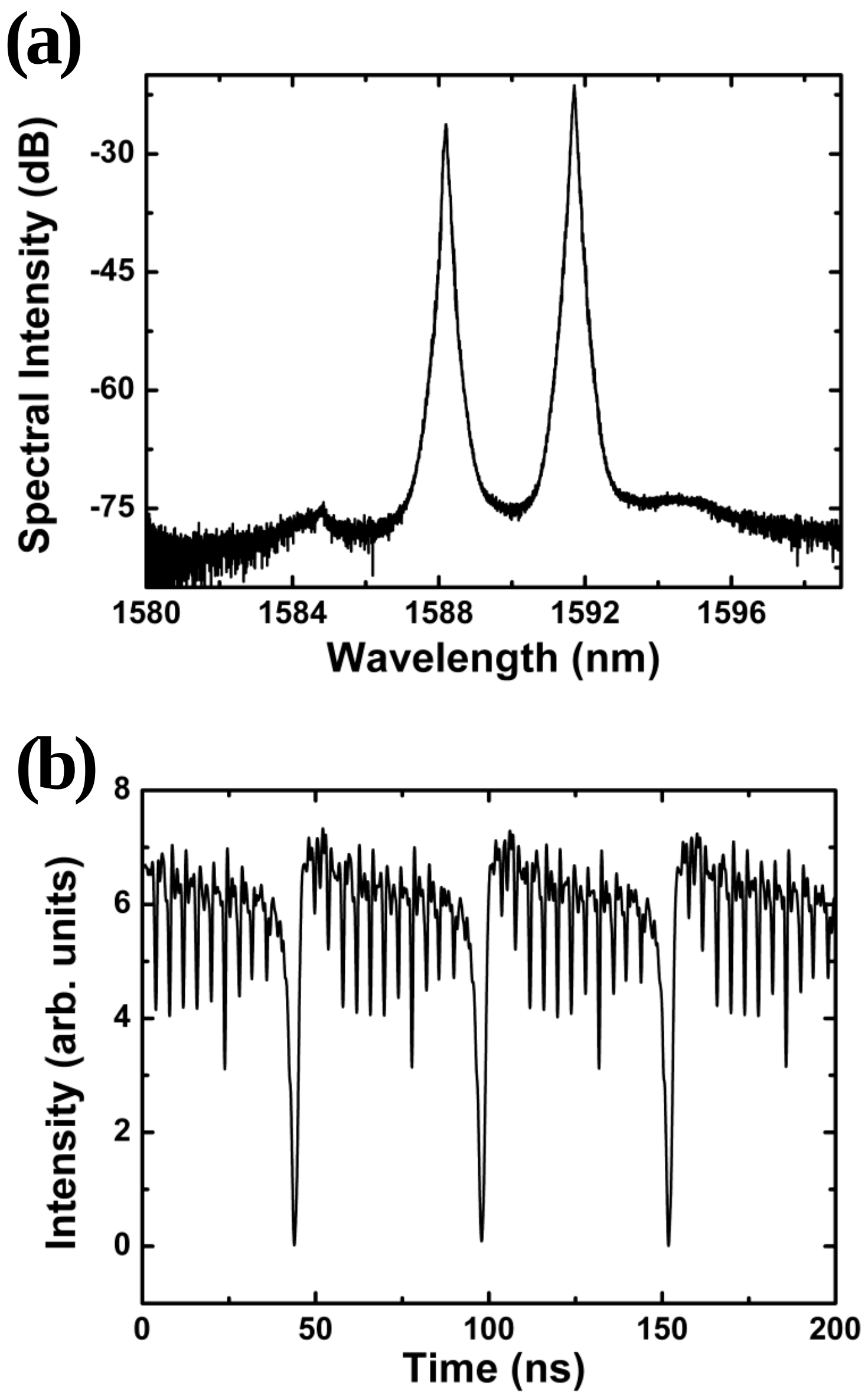

Fig.2 H. Zhang et. al.

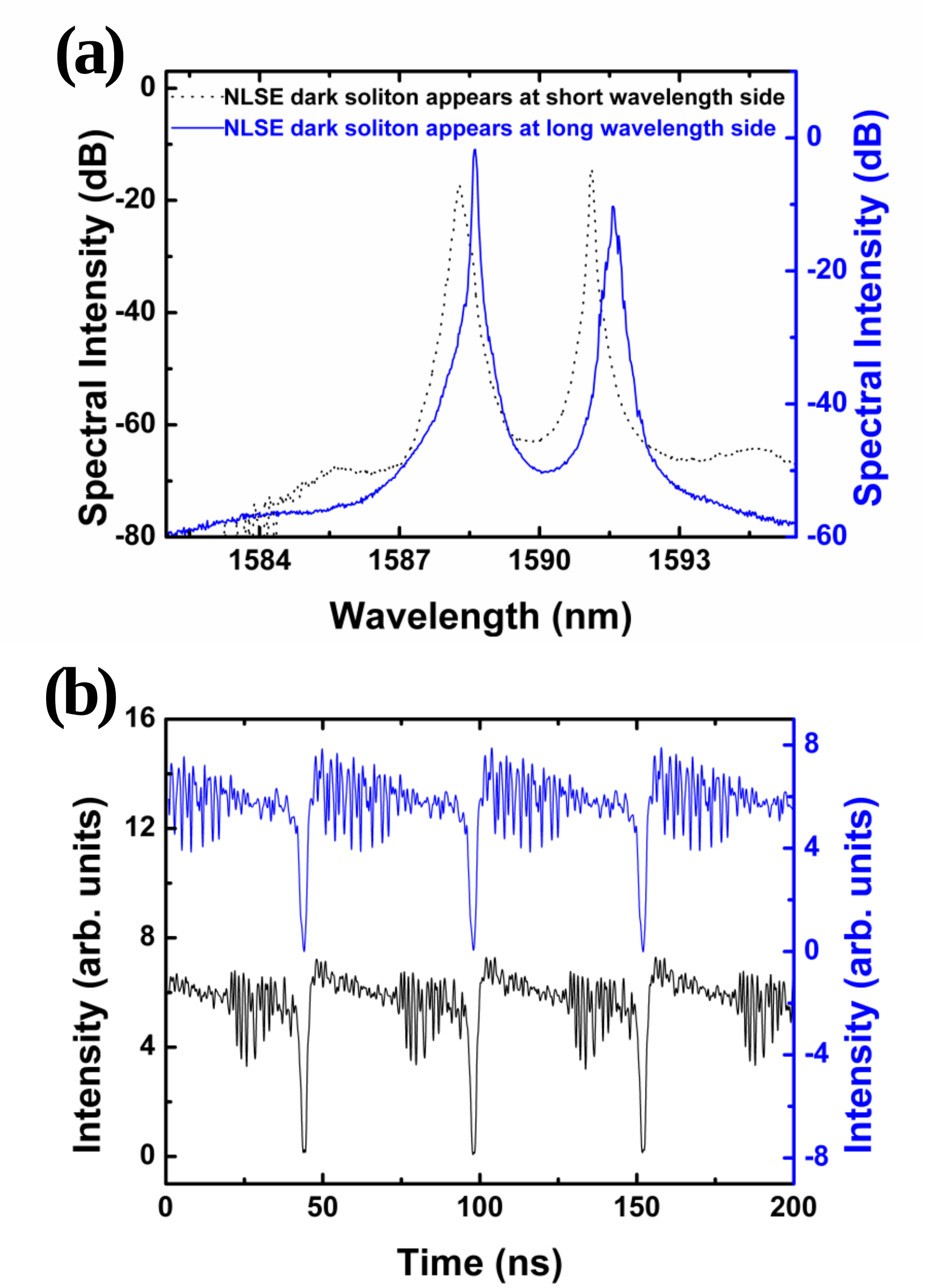

Fig.3 H. Zhang et. al.

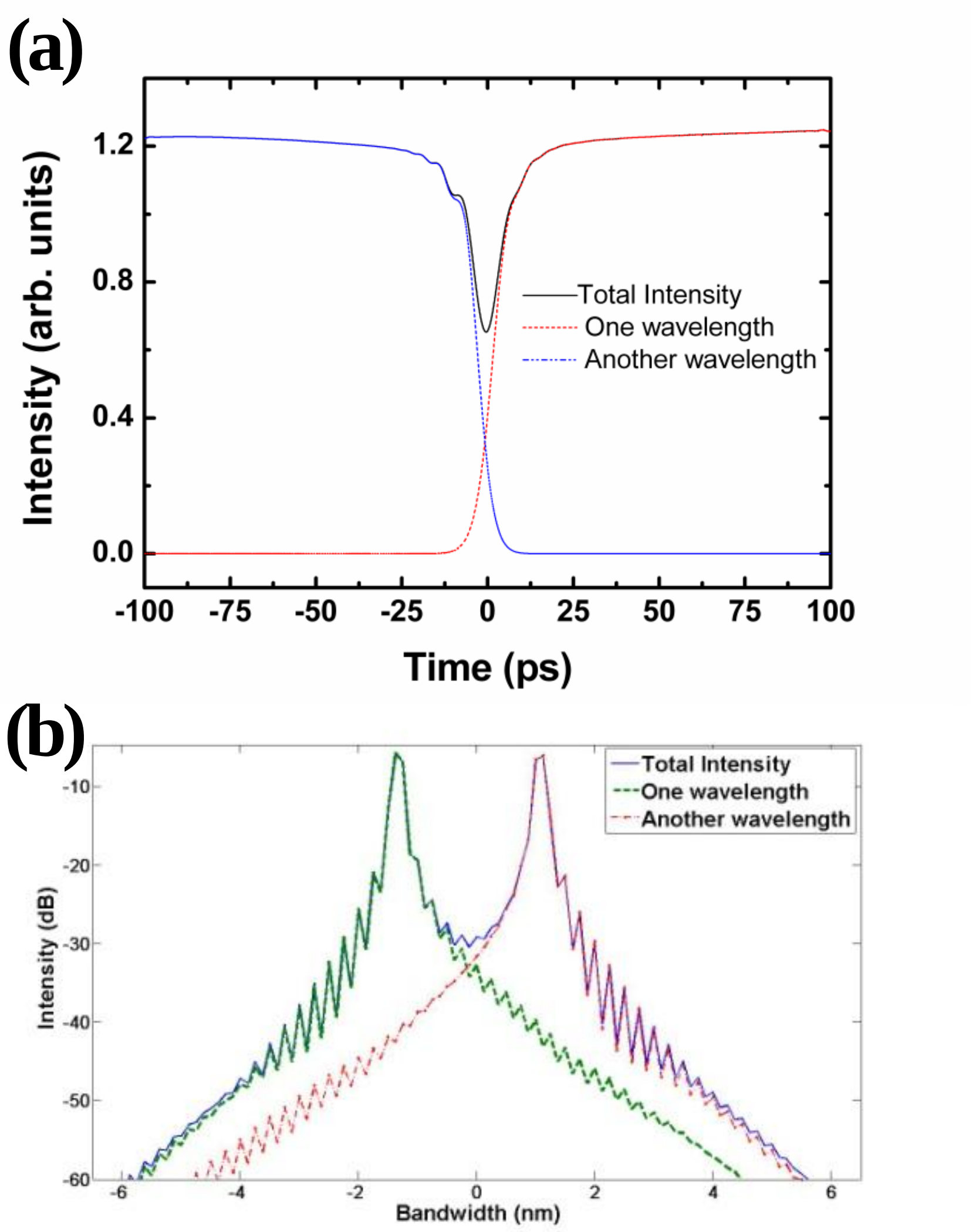

Fig.4 H. Zhang et. al.

## Supporting Figures for review purpose:

Fig. S1: Schematic of the dual wavelength domain wall fiber laser. WDM: wavelength division multiplexer. EDF: erbium doped fiber. PDI: polarization dependent isolator. PCs: polarization controllers.

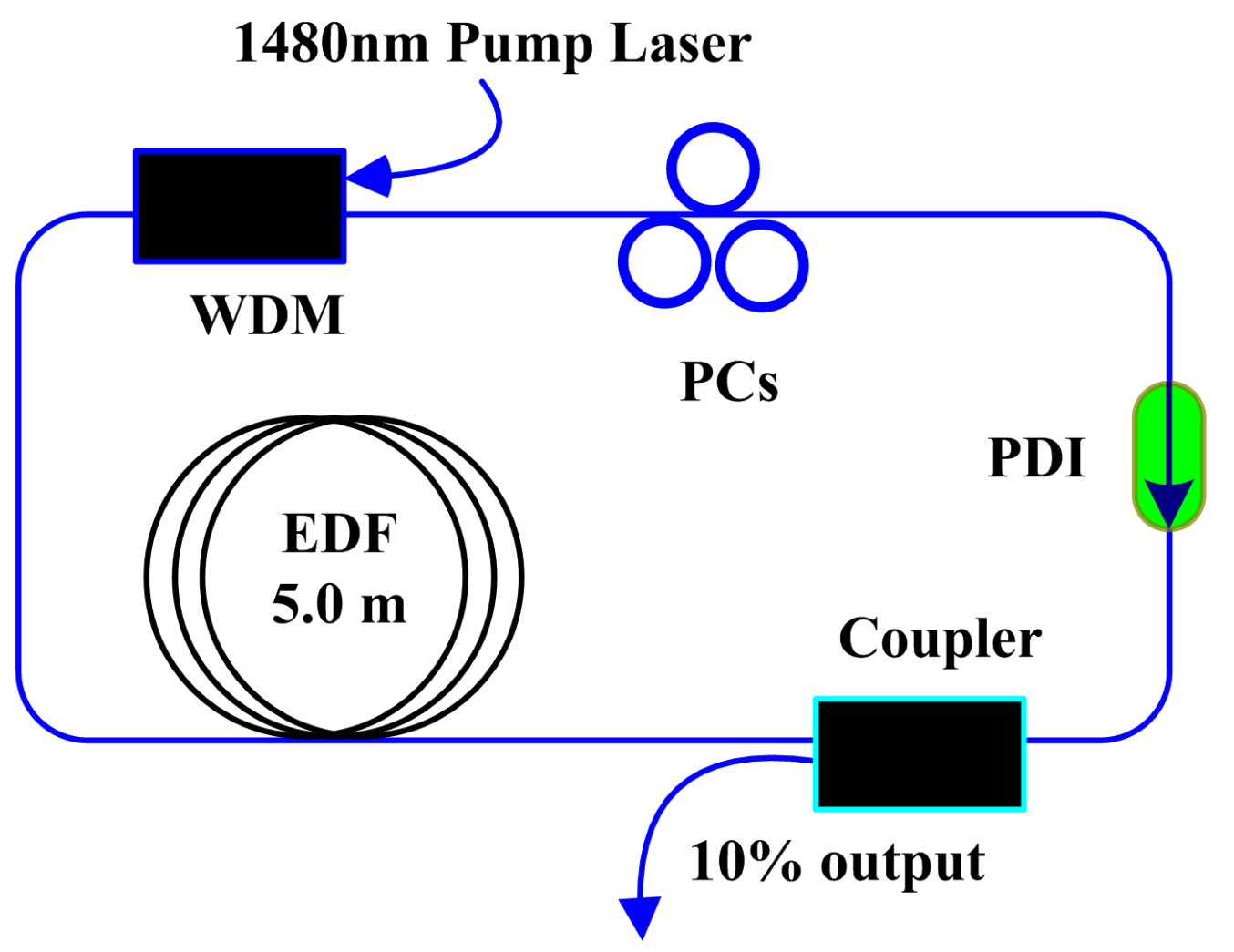

Fig. S1: H. Zhang et. al.

## Supporting Figures for review purpose:

Fig. S2: Generation of the NLSE type dark solitons at the same cavity but required much higher pump power. (a) Spectra and (b) oscilloscope traces.

(a)

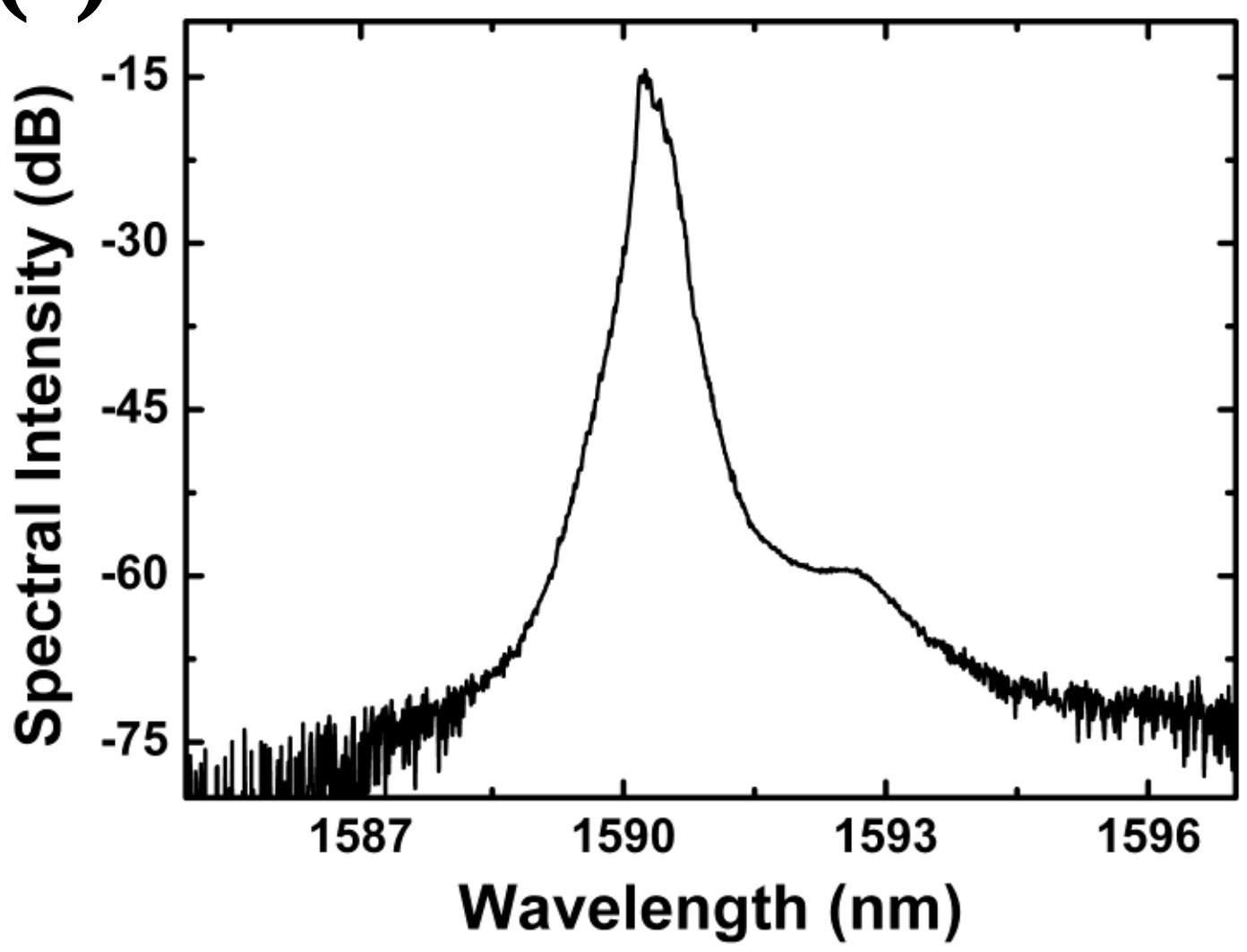

(b)

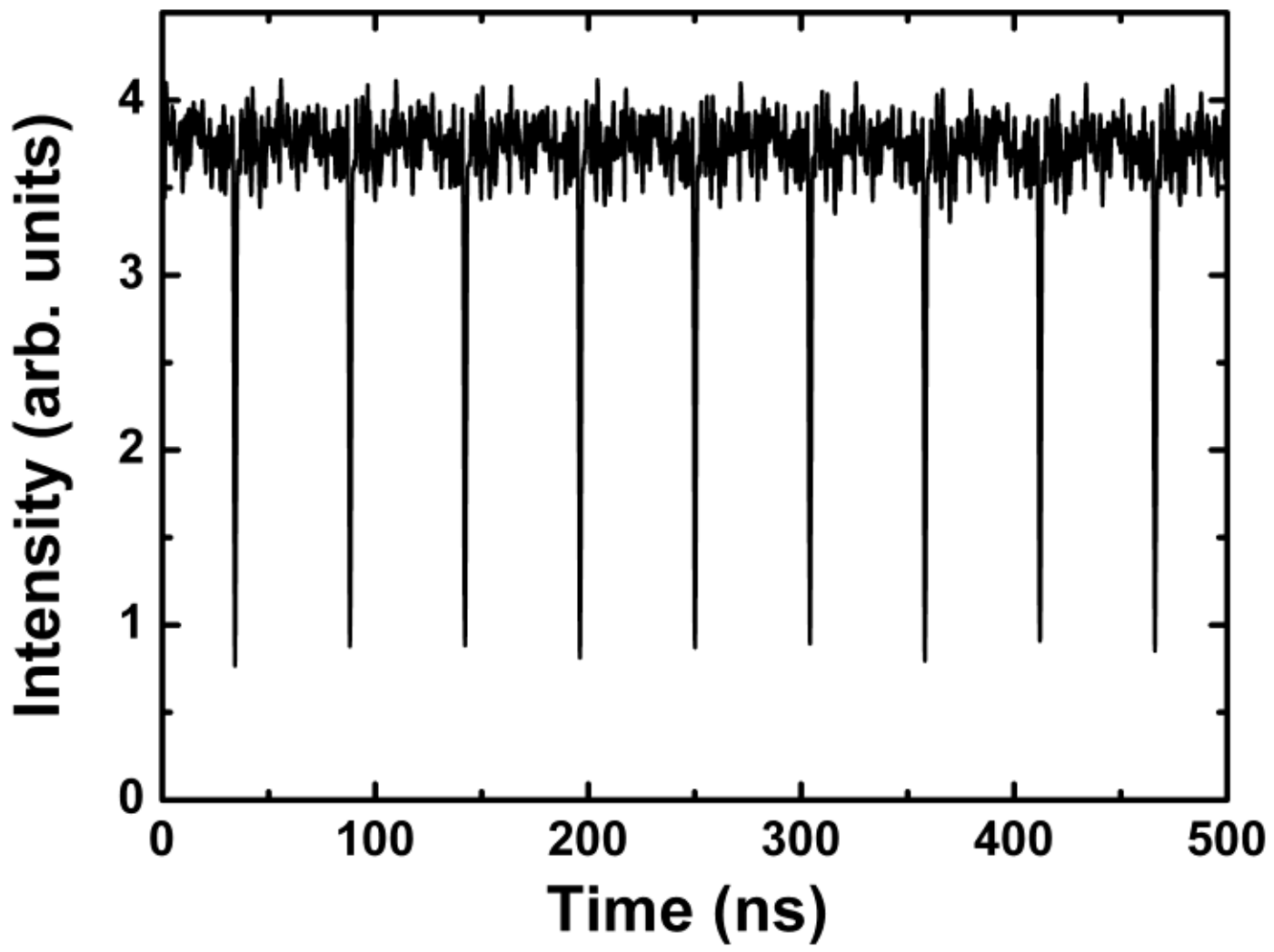

Fig. S2: H. Zhang et. al.

## Supporting Figures for review purpose:

Fig. S3: dual wavelength domain wall numerically calculated. Evolution of the dual wavelength domain wall with the cavity roundtrips: (a) one wavelength (shorter wavelength) (b) Another wavelength (longer wavelength).

**(a)**

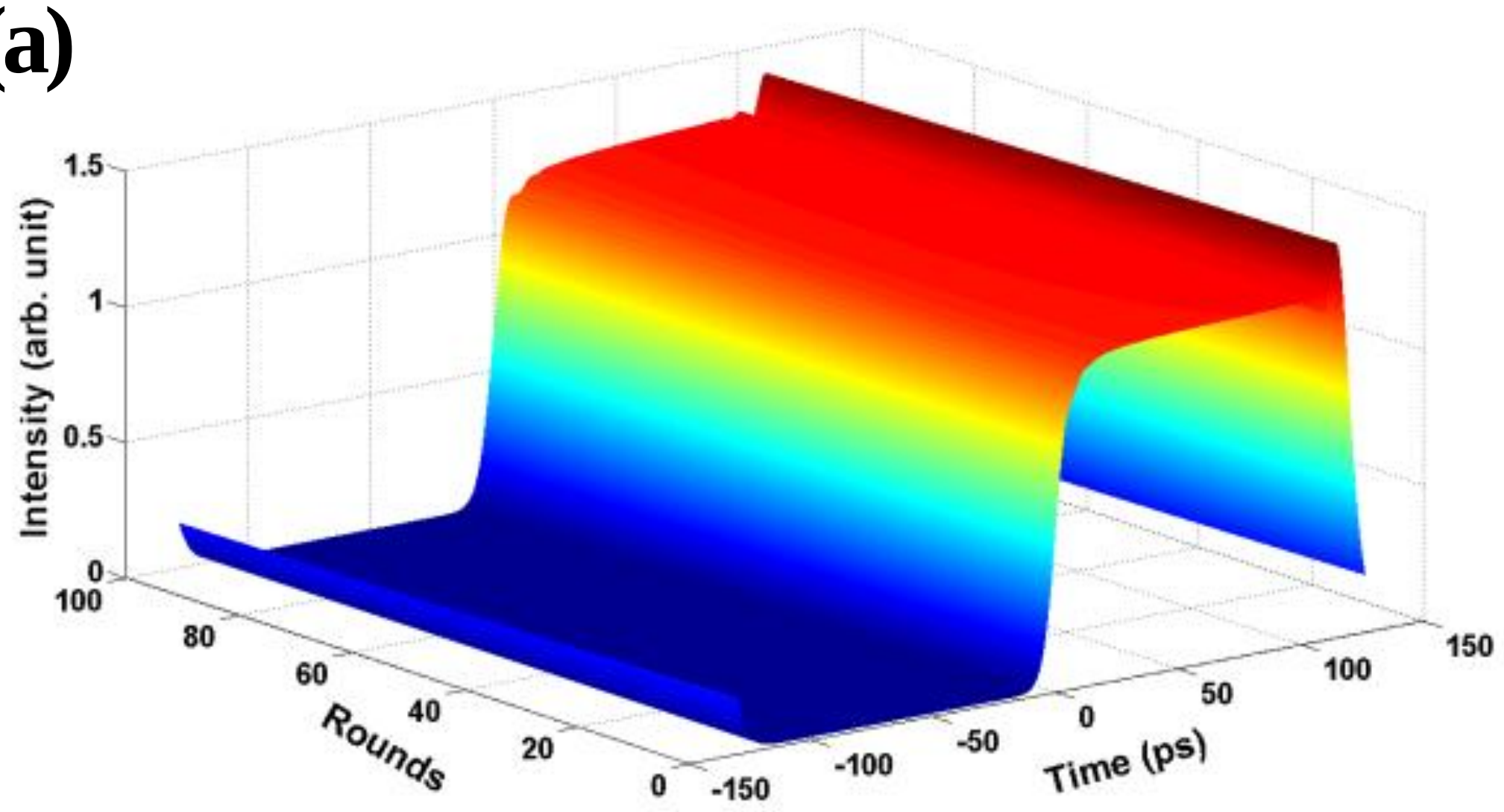

**(b)**

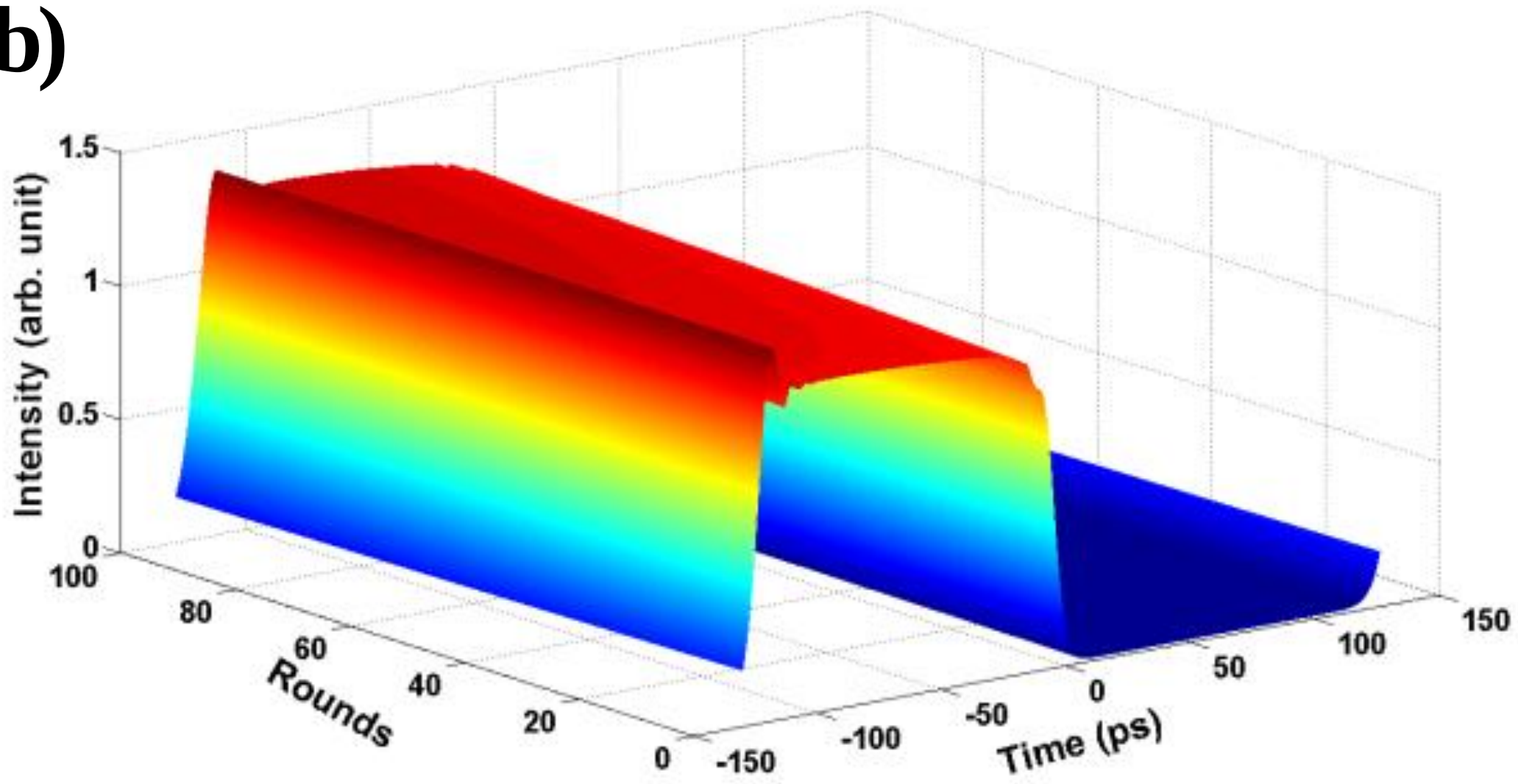

Fig. S3: H. Zhang et. al.